\documentstyle[11pt,epsfig]{article}
\begin{document}

\title{{\bf Stabilization of test particles in Induced Matter Kaluza-Klein
theory}}
\author{S. Jalalzadeh$^1$\thanks{%
email: s-jalalzadeh@sbu.ac.ir}, B. Vakili$^1$\thanks{%
email: b-vakili@sbu.ac.ir}, F. Ahmadi$^1$\thanks{%
email: fa-ahmadi@sbu.ac.ir} and H. R. Sepangi$^{1,2}$\thanks{%
email: hr-sepangi@sbu.ac.ir} \\
$^1${\small Department of Physics, Shahid Beheshti University, Evin, Tehran
19839, Iran}\\
$^2${\small Institute for Studies in Theoretical Physics and Mathematics,
P.O. Box 19395-5746, Tehran, Iran }}
\maketitle

\begin{abstract}
The stability conditions for the motion of classical test particles in an $%
n $-dimensional Induced Matter Kaluza-Klein theory is studied. We
show that stabilization requires a variance of the strong energy
condition for the induced matter to hold and that it is related to
the hierarchy problem. Stabilization of test particles in a FRW
universe is also discussed. \vspace{5mm}\newline PACS numbers:
04.20.-q, 040.50.+h, 040.60.-m
\end{abstract}

\vspace{2cm}

\section{Introduction}

Recent interests in Einstein theory of general relativity in five
and higher dimensions have been remarkable. Higher dimensional
extensions to general relativity were originally started with the
work of Kaluza and Klein (KK) with the addition of one extra
dimension and subsequently generalized to more extra dimensions by
various authors \cite{1,2}. Since the observable universe seems to
be 4-dimensional, these extra dimensions are assumed to have
compact topology with certain compactification scale $L$ so that
at scales much larger than $L$, such dimensions would not be
observable. If they exist, we may detect them at very short
distances or high energies. Another way to display the
invisibility of extra dimensions is to assume that the standard
matter is confined to a brane embedded in a higher dimensional
bulk manifold while the extra dimensions may only be probed by
gravitons. Such scenarios may be collectively termed as brane
theory.

In this paper we focus attention on the Induced Matter (IM) theory
\cite{3}. This theory is different from the classical KK scenario
by the fact that it has a noncompact fifth dimension and that the
$5D$ bulk space is devoid of matter. For this reason it is called
induced matter theory where the effective $4D$ matter is a
consequence of the geometry of the bulk. That is, in IM theory,
the $5D$ bulk space is Ricci-flat while the $4D$ hypersurface is
curved by the $4D$ induced matter. We will extend the IM ideas to
$nD$ bulk and ask the following question: if the world has more
than four dimensions and is described by a theory like general
relativity why do we live in a $4D$ universe and what conditions
restrict us to experience physics in $4D$? In \cite{4} the authors
discussed the general conditions for stabilization of test
particles in a brane world scenario with an arbitrary dimensional
bulk space that satisfy Einstein field equations. In brane models
the matter fields are confined to a fixed brane via a delta
function or confining potential \cite{4}. In IM theories however,
the matter fields are not confined to a fixed hypersurface and
hence test particles, for example, can propagate in the bulk space
and live on a fixed brane. We will show that if the
energy-momentum tensor of the matter field satisfies what we call
the Machian Strong Energy Condition (MSEC) on the four velocity of
a test particle and the number of extra dimensions is greater than
one, then the test particle is stabilized about our fixed brane
and would not disappear.

The organization of paper is as follows: in section 2 we give a
brief review of the geometrical definitions. In section 3 we
describe the dynamics of test particles. Section 4 deals with the
stabilization conditions of such particles and in Section 5 we
present an example for the stabilization of test particles in a
FRW universe. Conclusions are drawn in the last section.

\section{Geometrical setup}

Consider the background manifold $\overline{V}_{4}$ isometrically embedded
in ${V}_{n}$ by a map ${\cal Y}:\overline{V}_{4}\rightarrow V_{n}$ such that
\begin{equation}
{\cal G}_{AB}{\cal Y}_{\,\,\,,\mu }^{A}{\cal Y}_{\,\,\,,\nu }^{B}=\bar{g}%
_{\mu \nu },\hspace{0.5cm}{\cal G}_{AB}{\cal Y}_{\,\,\ ,\mu }^{A}{\cal N}%
_{\,\,\ a}^{B}=0,\hspace{0.5cm}{\cal G}_{AB}{\cal N}_{\,\,\ a}^{A}{\cal N}%
_{\,\,\ b}^{B}=g_{ab}=\varepsilon _{a};\hspace{0.5cm}\varepsilon _{a}=\pm 1
\label{mama41}
\end{equation}%
where ${\cal G}_{AB}$ $(\bar{g}_{\mu \nu })$ is the metric of the bulk
(brane) space $V_{n}(\overline{V}_{4})$ in an arbitrary coordinate, $\{{\cal %
Y}^{A}\}$ $(\{x^{\mu }\})$ are the basis of the bulk (brane) and ${\cal N}%
_{a}^{A}$ are $(n-4)$ normal unit vectors orthogonal to the brane.
Perturbation of $\overline{V}_{4}$ in a sufficiently small neighborhood of
the brane along an arbitrary transverse direction $\zeta $ is given by
\begin{equation}
{\cal Z}^{A}(x^{\mu },\xi ^{a})={\cal Y}^{A}+({\cal L}_{\zeta }{\cal Y})^{A},
\label{B}
\end{equation}%
where ${\cal L}$ represents the Lie derivative. By choosing $\zeta ^{\mu }$
orthogonal to the brane we ensure gauge independency \cite{5} and have
perturbations of the embedding along a single orthogonal extra direction $%
\bar{{\cal N}}_{a}$, giving the local coordinates of the perturbed brane as
\begin{equation}
{\cal Z}_{,\mu }^{A}(x^{\nu },\xi ^{a})={\cal Y}_{,\mu }^{A}+\xi ^{a}\bar{%
{\cal N}}_{\,\,\ a,\mu }^{A}(x^{\nu }),  \label{C}
\end{equation}%
where $\xi ^{a}(a=5,...,n)$ are small parameters along ${\cal
N}_{a}^{A}$ parameterizing the extra noncompact dimensions. Also,
one can see from equation (\ref{B}) that since the vectors
$\bar{{\cal N}}^{A}$ depend only on the local coordinates $x^{\mu
}$, they do not propagate along the extra dimensions
\begin{equation}
{\cal N}_{\,\,\ a}^{A}(x^{\mu })=\bar{{\cal N}}_{\,\,\ a}^{A}+\xi ^{b}[\bar{%
{\cal N}}_{b},\bar{{\cal N}}_{a}]^{A}=\bar{{\cal N}}_{\,\,\ a}^{A}.
\label{D}
\end{equation}%
The above assumptions lead to the embedding equations of the perturbed
geometry
\begin{equation}
{\cal G}_{\mu \nu }={\cal G}_{AB}{\cal Z}_{\,\,\ ,\mu }^{A}{\cal Z}_{\,\,\
,\nu }^{B},\hspace{0.5cm}{\cal G}_{\mu a}={\cal G}_{AB}{\cal Z}_{\,\,\ ,\mu
}^{A}{\cal N}_{\,\,\ a}^{B},\hspace{0.5cm}{\cal G}_{AB}{\cal N}_{\,\,\ a}^{A}%
{\cal N}_{\,\,\ b}^{B}={\cal G}_{ab}.  \label{E}
\end{equation}%
If we set ${\cal N}_{\,\,\ a}^{A}=\delta _{a}^{A}$, the metric of the bulk
space can be written in the following matrix form (Gaussian frame)
\begin{equation}
{\cal G}_{AB}=\left( \!\!\!%
\begin{array}{cc}
g_{\mu \nu }+A_{\mu c}A_{\,\,\nu }^{c} & A_{\mu a} \\
A_{\nu b} & g_{ab}%
\end{array}%
\!\!\!\right) ,  \label{F}
\end{equation}%
where
\begin{equation}
g_{\mu \nu }=\bar{g}_{\mu \nu }-2\xi ^{a}\bar{K}_{\mu \nu a}+\xi ^{a}\xi ^{b}%
\bar{g}^{\alpha \beta }\bar{K}_{\mu \alpha a}\bar{K}_{\nu \beta b},
\label{G}
\end{equation}%
is the metric of the perturbed brane, so that
\begin{equation}
\bar{K}_{\mu \nu a}=-{\cal G}_{AB}{\cal Y}_{\,\,\,,\mu }^{A}{\cal N}_{\,\,\
a;\nu }^{B},  \label{H}
\end{equation}%
represents the extrinsic curvature of the original brane (second
fundamental form). Also, we use the notation $A_{\mu c}=\xi
^{d}A_{\mu cd}$, where
\begin{equation}
A_{\mu cd}={\cal G}_{AB}{\cal N}_{\,\,\ d;\mu }^{A}{\cal N}_{\,\,\ c}^{B}=%
\bar{A}_{\mu cd},  \label{I}
\end{equation}%
represent the twisting vector fields (normal fundamental form). Any fixed $%
\xi ^{a}$ signifies a new perturbed geometry, enabling us to define an
extrinsic curvature similar to the original one by
\begin{equation}
\widetilde{K}_{\mu \nu a}=-{\cal G}_{AB}{\cal Z}_{\,\,\ ,\mu }^{A}{\cal N}%
_{\,\,\ a;\nu }^{B}=\bar{K}_{\mu \nu a}-\xi ^{b}\left( \bar{K}_{\mu \gamma a}%
\bar{K}_{\,\,\ \nu b}^{\gamma }+A_{\mu ca}A_{\,\,\ b\nu }^{c}\right) .
\label{J}
\end{equation}%
Note that definitions (\ref{F}), (\ref{G}) and (\ref{J}) require
\begin{equation}
\widetilde{K}_{\mu \nu a}=-\frac{1}{2}\frac{\partial {\cal G}_{\mu \nu }}{%
\partial \xi ^{a}}.  \label{M}
\end{equation}%
In geometric language, the presence of gauge fields $A_{\mu a}$ tilts the
embedded family of sub-manifolds with respect to the normal vector ${\cal N}%
^{A}$. According to our construction, the original brane is orthogonal to
the normal vector ${\cal N}^{A}.$ However, equation (\ref{E}) shows that
this is not true for the deformed geometry. Let us change the embedding
coordinates and set
\begin{equation}
{\cal X}_{,\mu }^{A}={\cal Z}_{,\mu }^{A}-g^{ab}{\cal N}_{a}^{A}A_{b\mu }.
\label{mama40}
\end{equation}%
The coordinates ${\cal X}^{A}$ describe a new family of embedded
manifolds whose members are always orthogonal to ${\cal N}^{A}$.
In this coordinates,  the embedding equations of the perturbed
brane is similar to the original one  described by equations
(\ref{mama41}), so that ${\cal Y}^{A}$ is replaced by ${\cal
X}^{A}$. This new embedding of the local coordinates are suitable
for obtaining induced Einstein field equations on the brane. We
will return to this point later in section 4. The extrinsic
curvature of the perturbed brane then becomes
\begin{equation}
K_{\mu \nu a}=-{\cal G}_{AB}{\cal X}_{,\mu }^{A}{\cal N}_{a;\nu }^{B}=\bar{K}%
_{\mu \nu a}-\xi ^{b}\bar{K}_{\mu \gamma a}\bar{K}_{,\,\,\nu b}^{\gamma }=-%
\frac{1}{2}\frac{\partial g_{\mu \nu }}{\partial \xi ^{a}},
\label{mama42}
\end{equation}%
which is the generalized York relation and shows how the extrinsic curvature
propagates as a result of the propagation of the metric in the direction
normal to the original brane. In general, the new sub-manifold is an
embedding in such a way that the geometry and topology of the bulk space do
not become fixed \cite{5}. We now show that if the bulk space has certain
Killing vector fields then $A_{\mu ab}$ transform as the components of a
gauge vector field under the group of isometries of the bulk. Under a local
infinitesimal coordinate transformation for extra dimensions we have
\begin{equation}
\xi ^{\prime a}=\xi ^{a}+\eta ^{a}.  \label{N}
\end{equation}%
Assuming the coordinates of the brane are fixed, that is $x^{\prime \mu
}=x^{\mu }$ and defining
\begin{equation}
\eta ^{a}={\cal M}_{\,\,\ b}^{a}\xi ^{b},  \label{P}
\end{equation}%
then in the Gaussian coordinates of the bulk (\ref{F}) we have
\begin{equation}
g_{\mu a}^{\prime }=g_{\mu a}+g_{\mu b}\eta _{,a}^{b}+g_{ba}\eta
_{,\mu }^{b}+\eta ^{b}g_{\mu a,b}+{\cal O}(\xi ^{2}).  \label{O}
\end{equation}%
Hence the transformation of $A_{\mu ab}$ becomes
\begin{equation}
A_{\mu ab}^{\prime }=\frac{\partial g_{\mu a}^{\prime }}{\partial \xi
^{\prime b}}=\frac{\partial g_{\mu a}^{\prime }}{\partial \xi ^{b}}-\eta
_{,b}^{A}\frac{\partial g_{\mu a}^{\prime }}{\partial x^{A}}.  \label{S}
\end{equation}%
Now, using $\eta _{\,\,\ ,b}^{a}={\cal M}_{\,\,\ b}^{a}(x^{\mu })$ and $\eta
_{\,\,\ ,\mu }^{a}={\cal M}_{\,\,\ b,\mu }^{a}\xi ^{b}$ we obtain
\begin{equation}
A_{\mu ab}^{\prime }=A_{\mu ab}-2A_{\mu c[a}{\cal M}_{b]}^{c}+{\cal M}%
_{ab,\mu }.  \label{R}
\end{equation}%
This is exactly the gauge transformation of a Yang-Mills gauge potential. In
our model the gauge potential can only be present if the dimension of the
bulk space is equal to or greater than six $(n\geq 6)$, because the gauge
fields $A_{\mu ab}$ are antisymmetric under the exchange of extra coordinate
indices $a$ and $b$.

Recently there has been an interest in finding KK gravitational
models for which the bulk space is not only Ricci flat but also
Riemann flat (${\cal R}_{AB}=0={\cal R}_{ABCD}$) \cite{6}. The
vanishing of the Riemann tensor means that we have the analog of
Minkowski space-time in $n$-dimensions. In this case, the bulk is
empty but we have an observable $4$-dimensional curved universe
that contains the induced matter fields. Now, let the bulk space
be of Minkowskian type with signature $(p,q)$. The tangent space
of the brane has
signature $(1,3)$ implying that the orthogonal space has the isometry group $%
SO(p-1,q-3)$. Now, if ${\cal L}^{ab}$ are the Lie algebra
generators of the this group we have
\begin{equation}
\lbrack {\cal L}^{ab},{\cal L}^{cd}]=C_{pq}^{abcd}{\cal L}^{pq},  \label{Y}
\end{equation}%
where $C_{pq}^{abcd}$ are the Lie algebra structure constants defined by
\begin{equation}
C_{pq}^{abcd}=2\delta _{p}^{[b}g^{a][c}\delta _{q}^{d]}.  \label{W}
\end{equation}%
On the other hand if $F^{\mu \nu }=F_{ab}^{\mu \nu }{\cal L}^{ab}$ is to be
the curvature associated with the vector potential$A^{\mu }=A_{ab}^{\mu }%
{\cal L}^{ab}$, we have
\begin{equation}
F_{\mu \nu }=A_{\nu ,\mu }-A_{\mu ,\nu }+\frac{1}{2}[A_{\mu },A_{\nu }],
\label{X}
\end{equation}%
or in component form
\begin{equation}
F_{\mu \nu ab}=A_{\nu ,\mu ab}-A_{\mu ,\nu ab}+\frac{1}{2}%
C_{ab}^{mnpq}A_{\mu mn}A_{\nu pq}.  \label{V}
\end{equation}

\section{Test particle dynamics}
We start with the following action which is equivalent to the
usual action for a test particle
\begin{equation}
{\cal I}=\frac{1}{2}\int_{\lambda _{A}}^{\lambda _{B}}d\lambda \left[
e^{-1}(\lambda ){\cal G}_{AB}\dot{{\cal Z}^{A}}\dot{{\cal Z}^{B}}%
-M^{2}e(\lambda )\right] ,  \label{eq1}
\end{equation}%
where $\lambda $ is an arbitrary parameter on the worldline with endpoints $%
A $ and $B$, $e(\lambda )$, the \textquotedblleft\ einbein," is a
new independent function and $M$ is the particle mass in the bulk
space. Variation of the action functional with respect to $e$ and
${\cal Z}^{A}$ leads to
\begin{equation}
e=\frac{1}{M}\sqrt{{\cal G}_{AB}\frac{d{\cal Z}^{A}}{d\lambda }\frac{d{\cal Z%
}^{B}}{d\lambda }}=\frac{1}{M}\frac{dS}{d\lambda },  \label{eq2}
\end{equation}%
and
\begin{equation}
\frac{d\dot{{\cal Z}^{A}}}{d\lambda }+\bar{\Gamma}_{BC}^{A}\dot{{\cal Z}^{B}}%
\dot{{\cal Z}^{C}}=\frac{\dot{e}}{e}\dot{{\cal Z}^{A}},  \label{eq3}
\end{equation}%
where an overdot represents derivative with respect to $\lambda $ and $S$ is
an affine parameter along the $n$-dimensional path of the particle. To
understand the nature of the motion in $4D$ we must take $A=\mu $ in this
equation and project it onto our $4D$ brane. Also, with $A=a$ we obtain the
motion along the extra dimensions. For this decomposition we need the
Christoffel symbols of the bulk space. Use of ${\cal G}_{AB}$ and its
inverse
\begin{equation}
{\cal G}^{AB}=\left( \!\!\!%
\begin{array}{cc}
g^{\mu \nu } & -A^{\mu a} \\
-A^{\nu b} & g^{ab}+A_{\alpha }^{a}A^{\alpha b}%
\end{array}%
\!\!\!\right) ,  \label{b}
\end{equation}%
leads to
\begin{eqnarray}
\bar{\Gamma}_{\alpha \beta }^{\mu }&=&\Gamma _{\alpha \beta }^{\mu
}+\frac{1}{2}\{A_{\alpha c}F_{\beta }^{\,\,\,\mu c}+A_{\beta
c}F_{\alpha }^{\,\,\,\mu c}\}-\widetilde{K}_{\alpha \beta a}A^{\mu a}-\frac{1%
}{2}(A_{c\alpha }A_{\,\,\,a\beta }^{c}+A_{c\beta }A_{\,\,\,a\alpha
}^{c})A^{a\mu }  \nonumber\\
\bar{\Gamma}_{\alpha a}^{\mu }&=&-\widetilde{K}_{\,\,\,\,\alpha
a}^{\mu }-\frac{1}{2}F_{\,\,\,\,\alpha a}^{\mu
}-\frac{1}{2}(A_{ab\alpha
}A^{b\mu }+A_{ab}^{\,\,\,\mu }A_{\,\,\alpha }^{b}),  \nonumber \\
\bar{\Gamma}_{ab}^{\mu }&=&\bar{\Gamma}_{bc}^{a}=0,
\nonumber\\
\bar{\Gamma}_{\alpha \beta }^{a}&=&\frac{1}{2}\left\{ \nabla
_{\beta }A_{\,\,\,\,\alpha }^{a}+\nabla _{\alpha }A_{\,\,\,\,\beta
}^{a}+A^{a\mu }A_{\,\,\,\,\alpha }^{c}F_{c\mu \beta }+A^{a\mu
}A_{\,\,\,\,\beta }^{c}F_{c\mu \alpha }\right\} +\widetilde{K}%
_{\,\,\,\,\alpha \beta }^{a}+A_{\,\,\,\,\mu }^{a}A^{\mu b}\widetilde{K}%
_{\alpha \beta b}-  \nonumber \\
& &\frac{1}{2}A^{a\mu }A_{b\mu }\left( A_{c\alpha }A_{\,\,\,\beta
}^{cb}+A_{c\beta }A_{\,\,\,\alpha }^{cb}\right) ,  \nonumber \\
\bar{\Gamma}_{b\alpha }^{a}&=&-\frac{1}{2}A^{a\beta }F_{b\alpha
\beta }+A_{\,\,\,\,b\alpha }^{a}+A^{a\mu }\widetilde{K}_{\alpha \mu b}-\frac{%
1}{2}A^{a\mu }\left( A_{bc\mu }A_{\alpha }^{c}+A_{bc\alpha }A_{\mu
}^{c}\right), \nonumber
\end{eqnarray}%
where $\Gamma _{\alpha \beta }^{\mu }$ are the Christoffel symbols induced
on the perturbed brane. Substituting these relations into (\ref{eq3}), the
equations of motion split into the following forms
\begin{equation}
\frac{d^{2}x^{\mu }}{d\lambda ^{2}}+\Gamma _{\,\,\,\,\alpha \beta }^{\mu }%
\frac{dx^{\alpha }}{d\lambda }\frac{dx^{\beta }}{d\lambda }={\cal Q}%
^{a}F_{\,\,\,\,\alpha a}^{\mu }\frac{dx^{\alpha }}{d\lambda }%
+2K_{\,\,\,\,\alpha a}^{\mu }\frac{dx^{\alpha }}{d\lambda }\frac{d\xi ^{a}}{%
d\lambda }+K_{\alpha \beta a}A^{\mu a}\frac{dx^{\alpha }}{d\lambda }\frac{%
dx^{\beta }}{d\lambda }+\frac{\dot{e}}{e}\frac{dx^{\mu }}{d\lambda },
\label{d}
\end{equation}%
and
\begin{eqnarray}
\frac{d^{2}\xi ^{a}}{d\lambda ^{2}} &+&\left( \nabla _{\alpha }A_{\,\,\
\beta }^{a}+A^{a\mu }A_{\,\,\ \alpha }^{c}F_{c\mu \beta }+\widetilde{K}%
_{\,\,\ \alpha \beta }^{a}+A_{\,\,\ \mu }^{a}A^{\mu b}K_{\alpha \beta
b}\right) \frac{dx^{\alpha }}{d\lambda }\frac{dx^{\beta }}{d\lambda }
\nonumber  \label{e} \\
&+&\left( 2A_{\,\,\ b\alpha }^{a}-A^{a\beta }F_{b\alpha \beta }+2A^{a\mu
}K_{\alpha \mu b}\right) \frac{dx^{\alpha }}{d\lambda }\frac{d\xi ^{b}}{%
d\lambda }=\frac{\dot{e}}{e}\frac{d\xi ^{a}}{d\lambda },
\end{eqnarray}%
where
\begin{equation}
{\cal Q}^{a}=\frac{d\xi ^{a}}{d\lambda }+A_{\beta }^{a}\frac{dx^{\beta }}{%
d\lambda },  \label{f}
\end{equation}%
is a {\it charge}-like quantity (per unit mass) associated with the
particle. It is clear that this quantity appears because of the motion in
the extra dimensions. With this definition equation (\ref{e}) simplifies to
\begin{equation}
\frac{d{\cal Q}^{a}}{d\lambda }=-\frac{dx^{\alpha }}{d\lambda }\left(
{K}_{\,\,\,\,\alpha \beta }^{a}\frac{dx^{\beta }}{d\lambda }%
+A_{\,\,\,\,\alpha b}^{a}{\cal Q}^{b}\right) +\frac{\dot{e}}{e}{\cal Q}^{a}.
\label{g}
\end{equation}%
Using the explicit form of the line element of the bulk space
\begin{equation}
dS^{2}=ds^{2}+g_{ab}\left( d\xi ^{a}+A_{\mu }^{a}dx^{\mu }\right) \left(
d\xi ^{b}+A_{\nu }^{b}dx^{\nu }\right)  \label{eq4}
\end{equation}%
where $ds^{2}=g_{\mu \nu }dx^{\mu }dx^{\nu }$ is the line element of the
perturbed brane, equation (\ref{eq2}) gives
\begin{equation}
e^{2}=M^{-2}\left[ \left( \frac{ds}{d\lambda }\right) ^{2}+g_{ab}{\cal Q}^{a}%
{\cal Q}^{b}\right] .  \label{eq5}
\end{equation}%
Now, differentiating equation
\begin{equation}
\left( \frac{ds}{d\lambda }\right) ^{2}=g_{\mu \nu }\frac{dx^{\mu }}{%
d\lambda }\frac{dx^{\nu }}{d\lambda }=g_{\mu \nu }u^{\mu }u^{\nu },
\label{mama}
\end{equation}%
we obtain
\begin{equation}
u_{\mu }\left[ \dot{u}^{\mu }+\Gamma _{\alpha \beta }^{\mu
}u^{\alpha }u^{\beta }\right] =\dot{s}\ddot{s}+{K}_{\,\,\,\nu
a}^{\mu }u^{\nu }u^{a}u_{\mu }.  \label{mama1}
\end{equation}%
If we contract the equation of motion (\ref{d}) with $u_{\mu }$ we
find
\begin{equation}
u_{\mu }\left[ \dot{u}^{\mu }+\Gamma _{\alpha \beta }^{\mu }u^{\alpha
}u^{\beta }\right] =2K_{\,\,\,\alpha a}^{\mu }u^{\alpha }u^{a}u_{\mu
}+K_{\alpha \beta a}A^{\mu a}u^{\alpha }u^{\beta }u_{\mu }+\frac{\dot{e}}{e}%
u^{\mu }u_{\mu }.  \label{mama2}
\end{equation}%
Comparing equations (\ref{mama1}) and (\ref{mama2}) one has
\begin{equation}
\frac{\dot{e}}{e}u^{\mu }u_{\mu }=\dot{s}\ddot{s}-K_{\mu \nu a}{\cal Q}%
^{a}u^{\mu }u^{\nu }.
\end{equation}%
Now, assuming that $d\lambda =ds$, we then have $u^{\mu }u_{\mu }=1$ and
\begin{equation}
\frac{\dot{e}}{e}=-K_{\mu \nu a}{\cal Q}^{a}u^{\mu }u^{\nu }.  \label{mama3}
\end{equation}%
Inserting this relation into equation of motion (\ref{d}), we arrive at the
following form for the force per unit mass acting on the test particle
\begin{equation}
{\it f}^{\mu }={\cal Q}^{a}F_{\,\,\,\alpha a}^{\mu }u^{\alpha }+\left(
2g^{\mu \nu }-u^{\mu }u^{\nu }\right) {\cal Q}^{a}K_{a\alpha \nu }u^{\alpha
}+\left( K_{\alpha \beta a}A^{a\mu }-2K_{\,\,\,\alpha a}^{\mu
}A_{\,\,\,\beta }^{a}\right) u^{\alpha }u^{\beta }.  \label{eq7}
\end{equation}%
The existence of this force arises from the motion of the test
particle not only in $4D$ but also in the extra dimensions. The
first term in (\ref{eq7}) is a {\it Lorentz}-like force, the
existence of which is expected since we showed in section 2 that
the twisting vectors $A_{\mu ab}$ transform as gauge potentials in
Yang-Mills theory so that we may consider ${\cal Q}^{a}$ as the
ratio of charge to mass of the test particle
\begin{equation}
{\cal Q}^{a}=q^{a}/m,  \label{eq8}
\end{equation}%
where $q^{a}$ and $m$ are the observable mass and charge of the
particle respectively. Let us note that the orthogonal part of
${\it f}$ does not vanish here. This is an interesting result, for
all known basic $4D$ forces lead to accelerations that are
orthogonal to the $4D$ velocity of the particle. Some authors
\cite{7} relate this timelike acceleration to $4D$ physics by
assuming that the \textquotedblleft invariant\textquotedblright\
inertial mass of a test particle varies along its worldline and
the timelike acceleration corresponds to this variation of the
inertial mass. In what follows, we see how this idea can be
generalized.

To obtain the effective $4D$ observable mass $m$ that an observer
measures, consider Lagrangian (\ref{eq1}) from which the momentum
conjugate to ${\cal Z}^{A}$ is
\begin{equation}
P_{A}=\frac{\partial {\cal L}}{\partial \dot{{\cal Z}^{A}}}=e^{-1}{\cal G}%
_{AB}\dot{{\cal Z}}^{B},  \label{eq10}
\end{equation}%
so that from the line element (\ref{eq4}) we have
\begin{equation}
 {\cal G}^{AB}P_{A}P_{B}-M^{2} =0.
\end{equation}%
In order to obtain the induced mass and charge of test particles
we project the $n$-momenta $P^{A}$ into four dimensions. This
projection is done by the vielbeins ${\cal Z}^A_{,\mu}$
\cite{Isra}
\begin{equation}
p_\mu = {\cal G}_{AB}P^A{\cal Z}^B_{,\mu} = \frac{1}{e}\left(
u_\mu + {\cal Q}^c A_{\mu c} \right)= \frac{1}{e}\left( u_\mu +
\frac{q^c}{m} A_{\mu c}\right).\label{taze2}
\end{equation}
Now, a glance at the above equation shows that the choice
\begin{equation}
m=\frac{1}{e}, \label{eq11}
\end{equation}%
renders $p_\mu$ as the canonical momenta given by
$p_\mu=m(u_\mu+q^c A_{\mu c})$. Using the above relation and
equation (\ref{mama3}) we have
\begin{equation}
\frac{\dot{m}}{m} = -\frac{\dot{e}}{e} = {\cal
Q}^aK_{a\alpha\beta}u^\alpha u^\beta.\label{taze3}
\end{equation}
On the other hand, using (\ref{eq11}), (\ref{eq8}) and
(\ref{mama3}) we find the following differential equations for the
charge and mass of the particle
\begin{equation}
\dot{q}^{a}=-A_{\,\,\,\,b\alpha }^{a}q^{b}u^{\alpha
}-mK_{\,\,\,\,\alpha \beta }^{a}u^{\alpha }u^{\beta },
\label{eq12}
\end{equation}%
\begin{equation}
\dot{m}=q^{a}K_{a\alpha \beta }u^{\alpha }u^{\beta }.
\label{eq332}
\end{equation}%
The general solutions of these equations for $M\neq 0$ can be
written in the following form
\begin{equation}
\begin{array}{c}
m=M\cos \omega , \\
q^{a}=B^{a}\sin \omega ,%
\end{array}
\label{mama14}
\end{equation}%
where $B^{a}$ satisfies the differential equation
\begin{equation}
\dot{B}^{a}=-A_{b\beta }^{a}B^{b}u^{\beta },
\end{equation}%
so that relation (\ref{eq5}) implies $B^{a}B_{a}=M^{2}$ with
$\omega $ being the solution of the following differential
equation
\begin{equation}
\dot{\omega}=-\frac{1}{M}B^{a}K_{\alpha \beta a}u^{\alpha }u^{\beta }.
\end{equation}%
The above solutions will change when the mass of the particle
vanishes in the bulk space, that is $M=0$, leading to
\begin{equation}
\begin{array}{c}
q^{a}=B^{a}e^{\Omega }, \\
m=m_{0}e^{\Omega },%
\end{array}
\label{mama16}
\end{equation}%
where $m_{0}$ is the constant of integration satisfying $\
g_{ab}B^{a}B^{b}=-m_{0}^{2}$ \ and $\Omega $ and $B^{a}$ satisfy
\begin{eqnarray}
\dot{\Omega} =\frac{1}{m_{0}}B_{a}K_{\,\,\,\alpha \beta
}^{a}u^{\alpha
}u^{\beta }, \\
\dot{B}^a + B^a\dot{\Omega} + A^a_{\,\,\,b\alpha}u^{\alpha}B^b +
m_{0}K^a_{\,\,\,\alpha\beta}u^{\alpha}u^{\beta}=0.
\end{eqnarray}%
Finally, from equations (\ref{eq7}) and (\ref{eq12}) we have
\begin{equation}
{\it f}_\mu u^\mu = K_{a\alpha\beta}u^\alpha u^\beta u^a = {\cal
Q}^aK_{a \alpha\beta}U^\alpha u^\beta - K_{a\alpha\beta}A^a_\mu
u^\alpha u^\beta u^\mu = \frac{\dot{m}}{m} +
\frac{Dq^a}{d\lambda}A_{a\mu}u^\mu,\label{taze4}
\end{equation}
where
\begin{equation}
\frac{Dq^a}{d\lambda} = \dot{q}^a + A^a_{\,\,\, b\alpha}q^b
u^\alpha.\label{taze5}
\end{equation}
Equation (\ref{taze4}) shows that manifestations of the extra
force, as viewed by the $4D$ observer, are embodied in the
variation of inertial mass and charge of the particle.

\section{Stabilization}
Generally speaking, in non-compact multidimensional theories of
gravity our ordinary space-time may be represented as a $4D$
sub-manifold locally and isometrically embedded in an $nD$ bulk
space. In these theories, we assume that the path of test
particles correspond to curves in $nD$ bulk manifolds, in contrast
to brane models where test particles are necessarily confined to
the $4D$ sub-manifold. The interesting question at this point
comes from the equation of motion represented by (\ref{e}) which
describes the motion of a particle along extra dimensions. This
equation shows that the test particle can not be solely confined
to the background brane $\bar{g}_{\mu \nu }$ but would propagate
along the extra dimensions too. It is now appropriate to consider
the state of the particle living in the original sub-manifold;
under what conditions would the particle move back and forth about
this sub-manifold. Within ordinary scales of energy we do not see
the disappearance of particles and hence may assume that the
particle fluctuates about our $4D$ space-time. This point can be
seen in the first approximation of $\xi ^{a}$ of the second
equation of motion (\ref{e}), where we have assumed that the
initial velocity of a test particle along the extra dimensions
vanishes on the non-perturbed brane
\begin{equation}
\frac{d^{2}\xi ^{a}}{ds^{2}}+(A_{\,\,\ \beta ;\alpha }^{a}+\widetilde{K}%
_{\,\,\ \alpha \beta }^{a})u^{\alpha }u^{\beta }=0.
\end{equation}%
Now, using equation (\ref{J}), the above equation simplifies to
\begin{equation}
\frac{d^{2}\xi ^{a}}{ds^{2}}+\left[ \left( A_{\,\,\,b\beta ;\alpha
}^{a}-A_{\alpha c}^{\,\,\,\,\,\,\,a}A_{\,\,\,\,b\beta }^{c}-\bar{K}%
_{\,\,\alpha \gamma }^{a}\bar{K}_{\,\,\,\,\beta b}^{\gamma }\right)
u^{\alpha }u^{\beta }\right] \xi ^{b}+\bar{K}_{\,\,\,\alpha \beta
}^{a}u^{\alpha }u^{\beta }=0.  \label{aa}
\end{equation}%

Before concentrating on stabilization, it would be necessary to
make some of the concepts to be used  more transparent and clear
in what follows. Let us then start by making a quick look at the
generalized $n$-dimensional IM theory. This would help us to grasp
the salient points of our discussion more easily. In this theory,
the motivation for assuming the existence of large extra
dimensions was to achieve the unification of matter and geometry,
{\it i.e.}, to obtain the properties of matter as a consequence of
extra dimensions. In the IM approach, Einstein equations in the
bulk are written in the form
\begin{equation}
{\cal R}_{AB}=0,  \label{eq56}
\end{equation}%
where ${\cal R}_{AB}$ is the Ricci tensor of the $nD$ bulk space.
To obtain the effective field equations in $4D$, let us start by
contracting the Gauss-Codazzi equations \cite{9}
\footnote{Eisenhats's  convention \cite{9} has been used in
defining the Riemann tensor.}
\begin{equation}
R_{\alpha \beta \gamma \delta }=2g^{ab}K_{a[\gamma \alpha }K_{b\beta ]\delta
}+{\cal R}_{ABCD}{\cal X}_{,\alpha }^{A}{\cal X}_{,\beta }^{B}{\cal X}%
_{,\gamma }^{c}{\cal X}_{,\delta }^{D}
\end{equation}%
and
\begin{equation}
2K_{a\mu \lbrack \nu ;\rho ]}=2g^{cd}A_{[\rho ca}K_{\mu ]\nu d}+{\cal R}%
_{ABCD}{\cal X}_{,\mu }^{A}{\cal N}_{a}^{B}{\cal X}_{,\nu }^{C}{\cal X}%
_{,\rho }^{D}.  \label{eq58}
\end{equation}%
where ${\cal R}_{ABCD}$ and $R_{\alpha \beta \gamma \delta }$ are the
Riemann curvature of the bulk and perturbed brane respectively. To obtain
Ricci tensor and Ricci scaler of the $4D$ brane we contract the Gauss
equation
\begin{equation}
R_{\mu \nu }=g^{cd}(g^{\alpha \beta }K_{\mu \alpha c}K_{\nu \beta
d}-K_{c}K_{\mu \nu d})+{\cal R}_{ABCD}{\cal N}_{a}^{A}{\cal X}_{,\mu }^{B}%
{\cal X}_{,\nu }^{C}{\cal N}_{b}^{D},  \label{eq59}
\end{equation}%
and
\begin{equation}
R={\cal R}+(K\circ K-K_{a}K^{a})-2g^{ab}{\cal R}_{AB}{\cal N}_{a}^{A}{\cal N}%
_{b}^{B}+g^{ad}g^{bc}{\cal R}_{ABCD}{\cal N}_{a}^{A}{\cal N}_{b}^{B}{\cal N}%
_{c}^{C}{\cal N}_{d}^{D},  \label{eq60}
\end{equation}%
where we have denoted $K\circ K\equiv K_{a\mu \nu }K^{a\mu \nu }$ and $%
K_{a}\equiv g^{\mu \nu }K_{a\mu \nu }$. In the Gaussian form of
the metric of the bulk space, the last term appearing on the right
hand side of equation (\ref{eq60}) vanishes. Using equations
(\ref{eq59}) and (\ref{eq60}) we obtain the following relation
between the Einstein tensors of the bulk and brane
\begin{equation}
G_{AB}{\cal X}_{\mu }^{A}{\cal X}_{\nu }^{B}=G_{\mu \nu }-Q_{\mu \nu }-g^{ab}%
{\cal R}_{AB}{\cal N}_{a}^{A}{\cal N}_{b}^{B}g_{\mu \nu }+g^{ab}{\cal R}%
_{ABCD}{\cal N}_{a}^{A}{\cal X}_{\mu }^{B}{\cal X}_{\nu }^{C}{\cal N}%
_{b}^{D},  \label{eq62}
\end{equation}%
where $G_{AB}$ and $G_{\mu \nu }$ are the Einstein tensors of the
bulk and brane respectively, and
\begin{equation}
Q_{\mu \nu }=g^{ab}(K_{a\mu }^{\,\,\,\,\,\,\gamma }K_{\gamma \nu
b}-K_{a}K_{\mu \nu b})-\frac{1}{2}(K\circ K-K_{a}K^{a})g_{\mu \nu }.
\label{mama15}
\end{equation}%
Now, decomposing the Riemann tensor of the bulk space into the Weyl and
Ricci tensors and Ricci scalar and using equation (\ref{eq56}), the Einstein
field equations induced on the brane become
\begin{equation}
G_{\mu \nu }=Q_{\mu \nu }-{\cal E}_{\mu \nu },  \label{eq63}
\end{equation}%
where ${\cal E}_{\mu \nu }=g^{ab}{\cal C}_{ABCD}{\cal X}%
_{,\mu }^{A}{\cal N}_{a}^{B}{\cal N}_{b}^{C}{\cal X}_{,\nu }^{D}$
is the electric part of
the Weyl Tensor ${\cal C}_{ABCD}$. Note that directly from definition of $%
Q_{\mu \nu }$ it follows that it is independently a conserved
quantity, that is $Q_{\,\,\,\,\,\,;\mu }^{\mu \nu }=0$. All of the
above quantities in equation (\ref{eq63}) are obtained on the
perturbed brane since, according to the second equation of motion
(\ref{d}), the matter can not exactly be confined to the original
non perturbed brane. Hence from a $4D$ point of view, the empty
$nD$ equations look like Einstein equations with induced matter.
The electric part of the Weyl tensor is well known from the brane
point of view. It describe a traceless matter, denoted by dark
radiation or Weyl matter. As was mentioned before, $Q_{\mu \nu }$
is a conserved quantity which, according to the spirit of the  IM
theory should be related to the ordinary matter as partly having a
geometrical origin. In the $5D$ bulk space, the author of
reference \cite{10} tried to relate this quantity to
the matter content of the universe via the following setup: from equation (%
\ref{eq56}), the trace of the Codazzi equation (\ref{eq58}) in
$5D$ gives the gravi-vector equation
\begin{equation}
W_{\,\,\,\,\beta ;\alpha }^{\alpha }=0,  \label{eq64}
\end{equation}%
where $W_{\alpha \beta }=K_{\alpha \beta }-g_{\alpha \beta }K$. Decomposing
it we have
\begin{equation}
W_{\alpha \beta }=-\frac{\kappa _{(5)}^{2}}{2}(-\lambda g_{\alpha \beta
}+T_{\alpha \beta }),  \label{eq65}
\end{equation}%
where $\kappa _{(5)}^{2}$ is a $5D$ gravitational constant, $\lambda $ is
the tension of the brane in five dimensions and $T_{\alpha \beta }$ is the
energy-momentum tensor of the standard matter. From the above equations we
get
\begin{equation}
K_{\mu \nu }=-\frac{1}{2}\kappa _{(5)}^{2}\left[ T_{\mu \nu }-\frac{1}{3}%
g_{\mu \nu }(T-\lambda )\right] .
\end{equation}%
Substituting this equation into (\ref{mama15}), we obtain
\begin{equation}
G_{\mu \nu }=-\Lambda g_{\mu \nu }+8\pi G_{N}T_{\mu \nu }+\epsilon \kappa
_{(5)}^{4}\Pi _{\mu \nu }-\epsilon {\cal E}_{\mu \nu },  \label{eq67}
\end{equation}%
where
\begin{equation}
\Lambda =\frac{\epsilon }{12}\lambda ^{2}\kappa _{(5)}^{4},
\end{equation}%
\begin{equation}
8\pi G_{N}=\frac{\epsilon }{6}\lambda \kappa _{(5)}^{4},
\end{equation}%
and
\begin{equation}
\Pi _{\mu \nu }=-\frac{1}{4}T_{\mu \alpha }T_{\,\,\nu }^{\alpha }+\frac{1}{12%
}TT_{\mu \nu }+\frac{1}{8}g_{\mu \nu }T_{\alpha \beta }T^{\alpha \beta }-%
\frac{1}{24}g_{\mu \nu }T^{2}.
\end{equation}%
It is now seen from equations (\ref{eq64}) and (\ref{eq65}) that
\begin{equation}
T_{\,\,\,\nu ;\mu }^{\mu }=0.  \label{eq71}
\end{equation}%
Also, $G_{\,\,\,\nu ;\mu }^{\mu }=0$ implies
\begin{equation}
{\cal E}_{\,\,\,\,\,\nu ;\mu }^{\mu }=\kappa _{(5)}^{4}\Pi _{\,\,\,\,\,\nu
;\mu }^{\mu }.  \label{eq72}
\end{equation}%
These equations, that is (\ref{eq67}), (\ref{eq71}) and
(\ref{eq72}) are similar to that of the SMS \cite{13} brane
gravity. However, it is easily seen that the above results are
contradictory. The quantity $Q_{\mu \nu }$ in equation
(\ref{eq63}) is conserved and so is $G_{\,\,\,\nu ;\mu }^{\mu }=0
$. Hence we have
\begin{equation}
{\cal E}_{\,\,\,\,\nu ;\mu }^{\mu }=0.
\end{equation}%
This equation must be satisfied in our model, but it contradicts equation (%
\ref{eq72}) and hence the validity of relation (\ref{eq65}), made
out of matter and extrinsic curvature, is questionable. On the
other hand, by contracting Codazzi equation (\ref{eq58}) in $nD$
bulk space, the generalized gravi-vector equation becomes
\begin{equation}
K_{\,\,\,\,\nu a;\mu }^{\mu }-K_{a,\nu }-(A_{\alpha ca}K_{\,\,\,\nu
}^{\alpha \,\,\,\,c}-A_{\nu ca}K^{c})+2g^{mn}{\cal R}_{ABCD}{\cal N}_{a}^{A}%
{\cal N}_{m}^{B}{\cal X}_{,\nu }^{C}{\cal N}_{n}^{D}=0.
\end{equation}%
From this equation we cannot define any suitable relationship like equation (%
\ref{eq65}) between extrinsic curvature and matter contained on the brane.
To define a suitable relation we use the following setup. In an arbitrary $n$%
-dimensional bulk space the Ricci equations are given by \cite{9}
\begin{equation}
F_{\mu \nu ab}=-2g^{\alpha \beta }K_{\mu \lbrack \alpha a}K_{\nu ]\beta b}-%
{\cal R}_{ABCD}{\cal N}_{a}^{A}{\cal N}_{b}^{B}{\cal X}_{,\mu }^{C}{\cal X}%
_{,\nu }^{D},
\end{equation}%
This equation shows how the gauge fields $A_{a\mu }$ or the corresponding field strength $%
F_{\alpha \beta a}$ can be derived from extrinsic curvature. Now,
using the decomposition of the Riemann tensor in terms of the Weyl
tensor, one may show that
\begin{equation}
{\cal R}_{ABCD}{\cal N}_{a}^{A}{\cal N}_{b}^{B}{\cal X}_{,\mu }^{C}{\cal X}%
_{,\nu }^{D}=0,  \label{eq76}
\end{equation}%
and hence the Ricci and Codazzi relations lead to
\begin{equation}
F_{\mu \nu ab}=Q_{\nu \mu ab}-Q_{\mu \nu ab},  \label{eq77}
\end{equation}%
where
\begin{equation}
Q_{\mu \nu ab}=K_{\mu \alpha a}K_{\,\,\,\nu b}^{\alpha }-K_{a}K_{\mu \nu b}-%
\frac{1}{2}g_{\mu \nu }\left( K^{\alpha \beta a}K_{\alpha \beta
b}-K_{a}K_{b}\right) .  \label{aa1}
\end{equation}%
The above considerations lead  to
\begin{equation}
Q_{\mu \nu ab}=-\frac{8\pi G_{N}}{(n-4)}T_{\mu \nu }g_{ab}-E_{\mu \nu ab}+%
\frac{1}{(n-4)}g_{ab}A_{\mu mn}A_{\nu }^{\,\,\,\,mn}+A_{\nu ;\mu ab}-A_{\mu
ca}A_{\,\,\,\nu b}^{c},  \label{aa2}
\end{equation}%
where
\begin{equation}
E_{\mu \nu ab}=\frac{1}{4\pi }\left[ g^{\alpha \beta }F_{\alpha \mu
a}F_{\beta \nu b}^{\,\,\,\,\,\,\,}-\frac{1}{4}g_{\mu \nu }F_{\alpha \beta
a}F_{b}^{\alpha \beta }\right] ,  \label{eq80}
\end{equation}%
and $G_{N}$ is the four dimensional gravitational constant.
Multiplication of equation (\ref{eq76}) and (\ref{eq77}) by
$g^{ab}$ results in the energy-momentum tensor
\begin{equation}
Q_{\mu \nu }=-8\pi G_{N}T_{\mu \nu }-\frac{1}{4\pi }\left[ g^{\alpha \beta
}F_{\alpha \mu a}F_{\beta \nu }^{\,\,\,\,\,\,a}-\frac{1}{4}g_{\mu \nu
}F_{\alpha \beta a}F^{\alpha \beta a}\right] .  \label{eq888}
\end{equation}%
This completes the description of the induced Einstein equation on the
perturbed brane.

Multiplication of equations (\ref{aa1}) and (\ref{aa2}) by $%
u^{\mu }u^{\nu }$ in the non-perturbed brane results in
\begin{eqnarray}
\left( \bar{K}_{\mu \gamma a}\bar{K}_{\nu b}^{\gamma }-A_{ab\nu ;\mu
}+A_{\mu ca}A_{b\nu }^{c}\right) u^{\mu }u^{\nu } &=&-\frac{8\pi G_{N}}{(n-4)%
}g_{ab}\bar{T}_{\mu \nu }u^{\mu }u^{\nu }-\frac{1}{2}\left( \bar{K}_{\alpha
\beta a}\bar{K}_{\,\,\,\,\,b}^{\alpha \beta }-\bar{K}_{a}\bar{K}_{b}\right) .
\nonumber \\
&&  \label{eq833}
\end{eqnarray}%
If we multiply equations (\ref{aa1}) and (\ref{aa2}) by $g^{\mu \nu }$, we
obtain the last term in equation (\ref{eq833}) and finally we find
\begin{equation}
(\bar{K}_{\mu \gamma a}\bar{K}_{\nu b}^{\gamma }-A_{ab\nu ;\mu }+A_{\mu
ca}A_{b\nu }^{c})u^{\mu }u^{\nu }=-\frac{8\pi G_{N}}{(n-4)}\left( \bar{T}%
_{\mu \nu }u^{\mu }u^{\nu }+\frac{1}{2}\bar{T}\right) +\frac{1}{2}D^{\mu
}A_{ab\mu },  \label{eq83}
\end{equation}%
where $D_{\nu }A_{\,\,\,ab}^{\mu }=A_{\,\,\,\,ab;\nu }^{\mu
}-A_{\nu ac}A_{\,\,\,b}^{\mu \,\,\,\,\,c}$ and $\bar{T}_{\mu \nu
}$ is the induced energy-momentum tensor on the original brane.
Hence if we assume that, as the generalized Coulomb gauge $D_{\mu
}A_{\,\,\,ab}^{\mu }=0$, then using equation (\ref{eq83}) in
equation (\ref{aa}) we find
\begin{equation}
\frac{d^{2}\xi ^{a}}{ds^{2}}+\frac{8\pi G_{N}}{(n-4)}\left( \bar{T}_{\mu \nu
}u^{\mu }u^{\nu }+\frac{1}{2}\bar{T}\right) \xi ^{a}+\bar{K}_{\,\,\alpha
\beta }^{a}u^{\alpha }u^{\beta }=0.  \label{eq84}
\end{equation}%
The last step in obtaining the equation of motion for $\xi ^{a}$
is done by expanding the 4-velocity $u^{\mu }$ and line element
$ds^{2}$ in the above equation in terms of the corresponding
quantities in the non perturbed brane. After some algebra we
obtain
\begin{equation}
\frac{d^{2}\xi ^{a}}{d\bar{s}^{2}}+\frac{8\pi G_{N}}{(n-4)}\left( \bar{T}%
_{\mu \nu }\bar{u}^{\mu }\bar{u}^{\nu }+\frac{1}{2}\bar{T}\right) \xi ^{a}+%
\bar{K}_{\,\,\alpha \beta }^{a}\bar{u}^{\alpha }\bar{u}^{\beta }=0,
\label{mama17}
\end{equation}%
where $d\bar{s}^{2}$ and $\bar{u}^{\mu }$ are the line element and
4-velocity
defined on the original brane respectively and the quantity $1/R_{a}=\bar{K}%
_{\,\,\alpha \beta }^{a}\bar{u}^{\alpha }\bar{u}^{\beta }$ is the
component of the normal curvature vector \cite{9}, its magnitude
being given by
\begin{equation}
\frac{1}{R^{2}}=|\bar{K}_{a\mu \nu }\bar{K}_{\,\,\alpha \beta }^{a}\bar{u}%
^{\mu }\bar{u}^{\nu }\bar{u}^{\alpha }\bar{u}^{\beta }|.  \label{mama18}
\end{equation}
In fact $1/R$ is nothing more than the higher dimensional
generalization of the familiar centripetal acceleration. Note that
in simplifying equation (\ref{mama17}) we have used the expansion
of the line element of the bulk space (\ref{G}), using the lowest
order approximation. This equation shows that the test particle
becomes stable around the original non perturbed brane if the
quantity inside the parenthesis is grater than zero. This, in turn
means that the energy-momentum tensor of the matter field
satisfies the restricted strong energy condition. As a
consequence, if the energy-momentum tensor of the matter field
vanishes, then according to equations (\ref{eq888}) and
(\ref{mama15}) the extrinsic
curvature vanishes. This means that according to equation (\ref%
{mama17}), the particle becomes totally unstable, its stability
becoming dependent on the existence of matter distribution over
the whole universe. Such a result seems to be in accordance with
the Mach's principle and for this reason we may call the above
energy condition as Machian.

At this point the question may arise as to even if the particle
motion is stabilized, we must make sure that the disappearance of
particles do not occur in our $4D$ universe. In what follows, we
show that in low energy physics, particles are gravitationally
trapped very near to our brane. The dynamical equation
(\ref{mama17}) shows that the potential energy acting on the
particle along the extra dimensions is
\begin{equation}
V = \frac{4\pi G_{N}}{(n-4)}\left( \bar{T} _{\mu \nu }\bar{u}^{\mu
}\bar{u}^{\nu }+\frac{1}{2}\bar{T}\right)g_{ab}\xi^a \xi^b +
\bar{K}_{\,\,\alpha \beta }^{a}\bar{u}^{\alpha }\bar{u}^{\beta
}\xi^a.\label{mama43}
\end{equation}
According to equation (\ref{eq332}), components of the normal
curvature vector are proportional to the variation of the rest
mass of particles. Observations shows that this variation is very
small. In addition, if we assume that the energy momentum tensor
of the matter content of the universe in the low energy regime is
proportional to the energy density $\rho_m$, then by neglecting
the last term in equation (\ref{mama17}), equation (\ref{mama43})
gives
\begin{equation}
L\equiv \left(g_{ab}\xi^a\xi^b\right)^{\frac{1}{2}} \sim
\left(\frac{V}{G_N\rho_m}\right)^{\frac{1}{2}},\label{mama44}
\end{equation}
where $L$ is the typical distanced traversed by the particle along
the extra dimensions. However, gravity is of multidimensional
nature in the bulk space and if we assume that the trapping of
particles is a gravitational effect, then stabilization caused by
the potential $V$ is equal to the gravitational energy of the
induced matter on the original non-perturbed brane, that is
\begin{equation}
V\sim \frac{G_{*}\rho_m}{L^{n-6}},\label{mama45}
\end{equation}
where $G_{*}$ is the fundamental $n$-dimensional Newton's constant
in the bulk space. Now, comparison of equations (\ref{mama44}) and
(\ref{mama45}) points to a similar result presented in \cite{nima}
without the product topology, namely
\begin{equation}
L^{n-4} \sim \frac{M^2_{Pl}}{M_{*}^{n-2}},\label{mama46}
\end{equation}
where $M_{*}=G_*^{-1/(n-2)}$ is the energy scale in the bulk
space. Assuming that $M_{*}\sim 1$ Tev, one calculates from
equation (\ref{mama46}) the value of $L$
\begin{eqnarray}
L \sim M_{*}^{-1}\left(\frac{M_{Pl}}{M_{*}}\right)^{\frac{2}{n-4}}
\sim 10^{\frac{32}{n-4}}\times 10^{-17}\mbox{cm}. \label{mama47}
\end{eqnarray}
For one extra dimension, the case commonly used in many IM models,
one obtains unacceptable values for $L$. In the case of two extra
dimensions,  we have that $L \sim 1 \hspace{1mm}\mbox{mm}$ and if
we assume $n=10$, as suggested by superstring theory, one has
$L\sim 10^{-12}\hspace{1mm}\mbox{cm}$. The result is that the
motion of test particles along the extra dimensions become very
limited and we can not see any disappearance of particles in our
$4D$ space-time at ordinary scales of energy.

Note that in this model, the size of the extra dimensions can be
very large. To see this, let us write the metric of the perturbed
brane as
\begin{equation}
g_{\mu \nu }=\bar{g}^{\alpha \beta }\left( \bar{g}_{\mu \alpha }-\xi ^{a}%
\bar{K}_{a\mu \alpha }\right) \left( \bar{g}_{\nu \beta }-\xi ^{a}\bar{K}%
_{a\nu \beta }\right) .  \label{mama19}
\end{equation}%
Since $\det{\cal G}=\det g_{\alpha \beta }\det g_{ab}\neq 0$, it follows that $%
\det g_{\alpha \beta }\neq 0$. Therefore, $\xi ^{a}$ cannot be a
solution of
\begin{equation}
\det\left( \bar{g}_{\alpha \beta }-\xi ^{a}\bar{K}_{a\alpha \beta
}\right) =0. \label{mama20}
\end{equation}%
The solutions of the above equation is defined as the curvature radii of the
background brane corresponding to each curvature line $dx^{\mu }$ and normal
${\cal N}^{a}$, satisfying
\begin{equation}
\left( \bar{g}_{\alpha \beta }-\xi ^{a}\bar{K}_{a\alpha \beta }\right)
dx^{\alpha }=0.  \label{mama21}
\end{equation}%
It follows that the metric of the perturbed brane becomes singular at the
solutions of equation (\ref{mama21}). Therefore, $g_{\mu \nu }$ and
consequently the metric of the bulk, become also singular at the points
determined by those solutions. Of course this is not a real singularity but
a property of the Gaussian system. However, this singularity breaks the
continuity and regularity of the Gauss-Codazzi-Ricci equations which are
constructed in this system. Therefore, the scale of curvature $\sigma $
defined by
\begin{equation}
\frac{1}{\sigma }=\sqrt{\bar{g}^{\mu \nu }g_{ab}\frac{1}{l_{a}^{\mu
}l_{b}^{\nu }}},  \label{mama23}
\end{equation}%
where $l_{a}^{\mu }$ is the curvature radii, sets a local limit
for the region in the bulk accessed by the gravitons. The behavior
of $\sigma $ near a space-time singularity depends on the
topological nature of that singularity. In the case of a
point-like singularity all values of $l_{a}^{\mu }$ tend to zero so that $%
\sigma $ also tends to zero. Therefore, the movement of particles
located near the singularity along the extra dimensions is very
limited and stability is again recovered. In the next section we
analyze a toy model to show how our formalism works.

\section{Stabilization of test particles in FRW universe}
In what follows, we will analyze the stabilization of test
particles in a FRW universe embedded in an $n$-dimensional flat
bulk space so that all the extra dimensions are spacelike. The FRW
line element (non perturbed brane) is written as
\begin{equation}
ds^{2}=-dt^{2}+a^{2}(t)\left[ dr^{2}+{\it f}^{2}(r)(d\theta ^{2}+\sin
^{2}\theta d\varphi ^{2})\right] ,  \label{eq86}
\end{equation}%
where ${\it f}(r)=\sin (r),r,\sinh (r)$ corresponds to $\kappa
=1,0,-1$ respectively and $a(t)$ is the scale factor. Now, to
construct the energy-momentum tensor of the induced matter fields,
we compute the components of the extrinsic curvature. From the
definition of extrinsic
curvature, equation (\ref{M}), it follows that for diagonal metric (\ref%
{eq86}), $K_{\mu \nu m}$ is also diagonal. In this case solving
the Codazzi equations (\ref{eq58}) with the assumption of
vanishing twisting vector fields, leads to
\begin{eqnarray}
K_{00m}=-\frac{1}{\dot{a}}\frac{d}{dt}\left( \frac{b_m}{a}\right) ,\hspace{0.5cm%
} \alpha,\beta = 0\nonumber\\
\\
K_{\alpha\beta m}=\frac{b_m}{a^{2}}g_{\alpha\beta}, \hspace{1.3
cm} \alpha,\beta \neq 0.\nonumber
\end{eqnarray}%
Here, $b_m(t)$ are arbitrary functions of $t$. Consequently, the
components of $Q_{\mu \nu }$ become
\begin{eqnarray}\label{eq188}
Q_{00}&=&-\frac{3}{a^{4}}g_{mn}b^mb^n,\nonumber\\
\\
Q_{\alpha\beta}&=&\frac{1}{a^4}\left( 2\frac{b_m\dot{b}^m}{H}-
b_mb^m\right) g_{\alpha\beta} \hspace{.5 cm}\alpha,\beta\neq
0,\nonumber
\end{eqnarray}
where $H=\frac{\dot{a}}{a}$ is the Hubble parameter. To proceed
with the geometrical interpretation of the energy-momentum tensor,
let us consider an analogy between $Q_{\mu \nu }$ and a simple
example of matter consisting of identical particles in the form of
dust, that is
\begin{equation}
T^{\mu \nu }=-\frac{1}{8\pi G_{N}}Q^{\mu \nu }=\rho u^{\mu }u^{\nu
}. \label{eq189}
\end{equation}
Using equations (\ref{eq188}) and (\ref{eq189}) and assuming that
the functions $b_m$ are equal, the energy density takes the
following form
\begin{equation}
\rho =\frac{3(n-4)}{8\pi G_{N}}\frac{b^{2}}{a^{4}}.\label{mama48}
\end{equation}
The $4D$ Einstein equations (\ref{eq63}), using equation
(\ref{eq188}), become
\begin{eqnarray}
& &\frac{\ddot{a}}{a}+2H^{2}+\frac{2k}{a^{2}}=\frac{(n-4)b}{\dot{a}%
a^{4}}\frac{d}{dt}(ab)  \nonumber \\
& & \\
& &\frac{\ddot{a}}{a}=\frac{(n-4)b^2}{a^{4}} -
(n-4)\frac{\dot{b}b}{H}. \nonumber
\end{eqnarray}
Elimination of $\ddot{a}$ from the above equations results in the Friedmann
equation
\begin{equation}
H^{2}+\frac{k}{a^{2}}=\frac{(n-4)b^{2}}{a^{4}}.
\end{equation}
Using equations (\ref{eq188}) and (\ref{eq189}) we also obtain
\begin{equation}
\frac{h}{H}=\frac{1}{2},\hspace{1cm}\bar{u}^{\mu }=-\delta _{0}^{\mu }.
\label{mama30}
\end{equation}
The solution of equation (\ref{mama30}) is
\begin{equation}
b=b_{0}\left( \frac{a}{a_{0}}\right) ^{1/2},  \label{mama31}
\end{equation}%
where $a_{0}$ is the present value of the scale factor and $b_{0}$
is an integration constant. Now the Friedmann equation becomes
\begin{equation}
H^{2}+\frac{k}{a^{2}}=\frac{8\pi G_{N}}{3}\rho.  \label{mama32}
\end{equation}
where
\begin{equation}
\rho=C\left(\frac{a}{a_0}\right)^{-3}, \hspace{.5 cm} C=
\frac{3(n-4)b_0^2}{8\pi G_N a_0^4}.
\end{equation}
Equation (\ref{mama32}) is exactly the standard Friedmann equation
in general relativity  for which one has textbook solutions. For a
closed universe we have $k=1$ and
\begin{equation}
\begin{array}{cc}
a=A\left( 1-\cos (\phi )\right) &  \\
t=A\left( \phi -\sin (\phi )\right) , &
\end{array}
\label{mama34}
\end{equation}%
where $A=b_{0}^{2}/2a_{0}$ and $\phi$ is the so-called development
angle. The geodesic equation (\ref{mama17}) then becomes
\begin{equation}
\frac{d^{2}\xi }{dt^{2}}+4\pi G_{N}\rho \xi +\left(8\pi G_{N}\rho
\right) ^{\frac{1}{2}}=0.  \label{mama35}
\end{equation}
It is also worth mentioning that according to equation (\ref{mama35}), for the cases when $k=0,\hspace{%
1mm} -1$, since the energy density $\rho$ approaches zero as the
time progresses, the particles become unstable on the brane.
However for a closed universe in the conformal time gauge,
equation (\ref{mama35}) changes to
\begin{equation}
\xi ^{a\prime \prime }-\frac{\sin(\phi )}{1-\cos(\phi )}\xi
^{a\prime }+\frac{3}{1-\cos(\phi )}\xi^a+\sqrt{2}(1-\cos(\phi
))^{\frac{1}{2}}=0,  \label{mama38}
\end{equation}
where a prime denotes derivation with respect to $\phi $. Figure 1
shows variation of $\xi^a$, obtained from equation (\ref{mama38}),
and the scale factor with respect to time from the big bang to the
big crunch.
\begin{figure}[tbp]
\centerline{\epsfig{figure=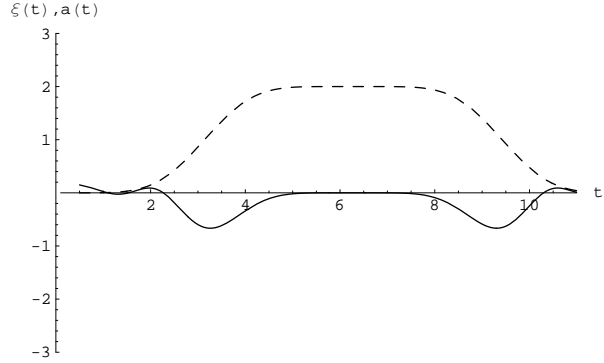,width=8cm}}
\caption{{\protect\footnotesize The behavior of the scale factor, dotted
line, and the extra dimension, solid line. The solid line shows that at the
present epoch (the plateau) a test particle is confined to the brane.}}
\label{fig1}
\end{figure}
An interesting comment at this stage relates to the size of extra
dimensions. In the previous section, we argued that the size of
extra dimensions are given by the curvature radii. If we compute
the scale of the curvature by using equations (\ref{eq188}),
(\ref{mama21}) and (\ref{mama23}) we find
\begin{equation}
\sigma \sim \frac{1}{\sqrt{G_{N}\rho}},
\end{equation}
showing that the average size of the extra dimensions is very
large. However, the multidimensional nature of gravity in the bulk
space, represented through equation (\ref{mama47}), forces the
particles becoming trapped near the brane.

\section{Conclusions}
We have studied a model in which a $4D$ brane is embedded in an $n$%
-dimensional bulk. In so far as a test particle moving in such a
bulk space is concerned, a $4D$ observer would detect a $4D$ mass
and Yang-Mills charges stemming from oscillations of the particle
along directions perpendicular to the brane if the confined matter
satisfies the Machian strong energy condition and the number of
extra dimensions is greater than one. A force is also experienced
by the particle which again originates from the effects of the
extra dimensions in the form of the extrinsic curvature and
Yang-Mills fields so that the terms arising from the extrinsic
curvature are rewritable as variations of the inertial mass and
charge of the test particle. An interesting point is that although
our model predicts extra dimensions which can have large sizes,
the multi-dimensional nature of gravity in the bulk space and its
energy scale forces a test particle to be confined to the brane or
very near it. This result should be viewed in the light of
5-dimensional IM theories commonly used in the literature where
the above stability conditions cannot be satisfied.

\end{document}